\documentclass[aps,pre,twocolumn,groupedaddress]{revtex4}

\usepackage{graphicx,color}

\begin{document}


\preprint{}

\title{Amplitude Modulation and Relaxation-Oscillation of 
Counterpropagating Rolls within a Broken-Symmetry Laser-Induced Electroconvection Strip}


\author{Daniel R. Spiegel}
\email[]{dspiegel@trinity.edu}
\affiliation{Department of Physics and Astronomy, Trinity University, San Antonio, Texas, USA}
\author{ Elliot R. Johnson}
\affiliation{Department of Physics and Astronomy, Trinity University, San Antonio, Texas, USA}
\author{Skyler R. Saucedo}
\affiliation{Department of Physics and Astronomy, Trinity University, San Antonio, Texas, USA}

\begin{abstract}
We report a liquid-crystal pattern-formation experiment in which we break the lateral (translational) symmetry of a nematic medium with a laser-induced thermal gradient.  The work is motivated by an improved measurement (reported here) of the temperature dependence of the electroconvection threshold voltage in planar-nematic 4-methoxybenzylidene-4-butylaniline (MBBA).  In contrast with other broken-symmetry-pattern studies that report a uniform drift, we observe a strip of counterpropagating rolls that collide at a sink point, and a strong temporally periodic amplitude modulation within a width of 3-4 rolls about the sink point.  The time dependence of the amplitude at a fixed position is periodic but displays a nonsinusoidal relaxation-oscillation profile.  After reporting experimental results based on spacetime contours and wavenumber profiles, along with a measurement of the change in the drift frequency with applied voltage at a fixed control parameter, we propose some potential guidelines for a theoretical model based on saddle-point solutions for Eckhaus-unstable states and coupled complex Ginzburg-Landau equations.  Published in PRE 73, 036317 (2006).

\end{abstract}

\date{\today}


\maketitle


\section{INTRODUCTION}
\label{sec-intro}
The formation of spatially periodic patterns in a wide variety of media driven far from equilibrium continues to attract a great deal of attention from experiment, simulations, and theory \cite{Peter-Zimmermann05, John05, Qui-Ahlers05, Boyer04,Buka04,Coullet04, Kamaga-DenninDislocations04,Toth-Katona-Gleeson04, Toth-Buka-Kramer02,Gheorghiu-Gleeson02,John02,Gollub-Langer99, CHReview}.  For patterns in fluids, many important  advances have relied on careful experiments with samples possessing a very high degree of spatial homogeneity in the material properties and the external control parameters.  This homogeneity has proved crucial for careful comparisons between experiments and theory, since the latter usually assumes that the medium has completely uniform spatial properties, resulting in a high degree of symmetry \cite{Ahlers89}.  Experimental patterns within samples of high purity and uniformity will undoubtedly continue to be of major importance in the foreseeable future.  On the other hand, patterns in nature almost never occur in highly homogenized media \cite{Kolodner92II}, and one may reasonably inquire whether an inhomogeneous fluid, with a symmetry that is broken due to the variation of either a material property or a control parameter (or both) along some specific direction, will give rise to significant new physics.  In the case of convection in horizontal fluid cells, broken up-down (U/D) symmetry about the midplane (non-Boussinesq convection) has been studied intensively for several decades and now forms a cornerstone of the Rayleigh-B\'{e}nard convection (RBC) problem \cite{Peter-Zimmermann05, CHReview}.  Although broken lateral symmetries (i.e., broken translational symmetry) in fluid media have received somewhat less attention than broken U/D symmetry, important results have been obtained.  A seminal work for broken lateral symmetry was the 1982 theoretical study by Kramer and coworkers \cite{Kramer82} on very small material-property ramps in pattern-forming media, which motivated a number of new experiments and simulations on ramped systems.  Hartung \emph{et al.} \cite{Hartung91}, for example, discovered convection patterns with a uniform group drift in a RBC experiment in which both U/D and lateral symmetry were broken due to the temperature-dependence of the viscosity and spatial modulations in the cell height.  Rehberg \emph{et al.} \cite{Rehberg-Busse87} showed that two ramps with opposite slope could force uniform phase drift in RBC.  Ning \emph{et al.} \cite{NingAhlersCannell90} found traveling vortex waves in a Taylor vortex experiment in which a ramp was created with an axial variation of the annular width.  Wiener and coworkers \cite{Wiener99} investigated ramped Taylor-vortex flow using an hourglass cell, which forms a particularly dramatic and intriguing example of broken lateral symmetry. Beyond fluids, there has also been considerable interest in spatially varying parameters in reaction-diffusion Turing systems \cite{Page05}.

It is the purpose of the present article to extend the understanding of pattern-formation in fluids with a broken lateral symmetry.  We report an experiment in which a controlled spatial ramp of the temperature within a nematic liquid crystal (NLC) produces spatially localized electroconvection rolls subjected to a positive nonuniform control parameter.  In a planar-aligned NLC, a preferred direction for orientation of the pattern is selected by the nematic director.  We create a long narrow strip-pattern parallel to the director, and break the lateral (translational) symmetry along the strip using a laser-driven horizontal thermal gradient.  The liquid crystal is planar-nematic 4-methoxybenzylidene-4-butylaniline (MBBA).  Our experiments, which expand on a preliminary study limited to much smaller pattern areas \cite{GiebinkDRS04}, rely on (1) the use of nonuniform laser irradiation to produce a long strip parallel to the director that is slightly warmer than its surroundings, and (2) the fact that the control parameter in this medium is (in general) strongly dependent on temperature, as we show directly in experimental detail below.  The experimental results are of interest because they differ qualitatively both from known near-threshold electroconvection patterns in planar MBBA  \cite{Rehberg-Ahlers91,RehbergAdvSolState89,Joets-RibottaPRL88}, and from work (cited above) on other systems with a broken lateral symmetry.  We find that the temporal variation of the roll amplitude at a fixed position is clearly nonsinusoidal.  We investigate the new dynamics by studying spacetime diagrams, the temporal variation of the amplitude at fixed positions, and wavenumber profiles.  Although a theory applicable to electroconvection in liquid crystals with strong spatial inhomogeneities is not available at present, we proffer suggestions, based on the experimental results, for potentially useful approaches to developing a theoretical model to explain our observations.  Many of these considerations should be applicable to other broken-symmetry pattern-forming fluids driven outside of equilibrium.

\section{Results and Discussion}
\label{sec-exper}


The experimental apparatus and samples employed for our measurements have been described previously \cite{GiebinkDRS04}.  Briefly, a nematic film of MBBA (Aldrich) is doped with a small amount of blue-absorbing dye (methyl red, MR) and placed within a conventional homemade electroconvection cell with a thickness of 40 $\mu \rm{m}$ and illuminated from above with a 488-nm laser beam.  The electroconvection pattern is viewed using an inverted microscope and a CCD camera with a 640 X 480 pixel array and a resolution of 0.54 pixels/$\mu \rm{m}$.   We choose spatial coordinates such that the $x$ and $y$ axes in the nematic plane are parallel and perpendicular to the nematic director, respectively.  For quantitative measurements of the pattern, we employ the background-divided dimensionless intensity, defined as \cite{Qui-Ahlers05,Bisang-Ahlers98}

\large
\begin{equation}
\label{EqIntensity} 
I(x,y,t) =  \frac{G(x,y,t)}{G_b(x,y)} - 1
\end{equation}
\normalsize
\\
where $G(x,y,t)$ and $G_b(x,y)$ are the sample and background gray-scale values, respectively, obtained at the pixel $(x,y)$ of the CCD array at time $t$.  The sample and background images are obtained under identical conditions except that for the latter, the voltage $V$ across the sample is set to zero.   At very low applied frequencies, we observed that the dimensionless intensity was spatially periodic along the director with a period equal to the sample thickness, as expected from shadowgraph theory \cite{TrainoffCannell02,RehbergJStatPhys91,RasenatRehberg89}.  The spatial period of the dimensionless intensity is thus one-half the spatial period of the director-tilt angle that defines the amplitude of the pattern.  The control parameter for an electroconvection experiment is defined as $\epsilon =  V^2/V_0^2 - 1$, where $V_0$ is the critical threshold voltage at which rolls first appear.  In our experiments, $\epsilon$ varies with position because of the non-uniform laser heating.  We consistently measure the threshold voltage $V_0$ at which rolls first appear along the laser-illuminated strip as the voltage is increased from zero.   As explained below, the rolls occur first at the point at which the laser intensity is largest.  Therefore, when the voltage across the sample is raised to $V > V_0$, the difference $V^2/V_0^2 - 1$ represents the maximum $\epsilon_{max}$ in the control parameter as a function of position.  We will employ $\epsilon_{max}$ defined in this way as our experimental control parameter.  

\begin{figure}
\includegraphics[width=0.55\textwidth]{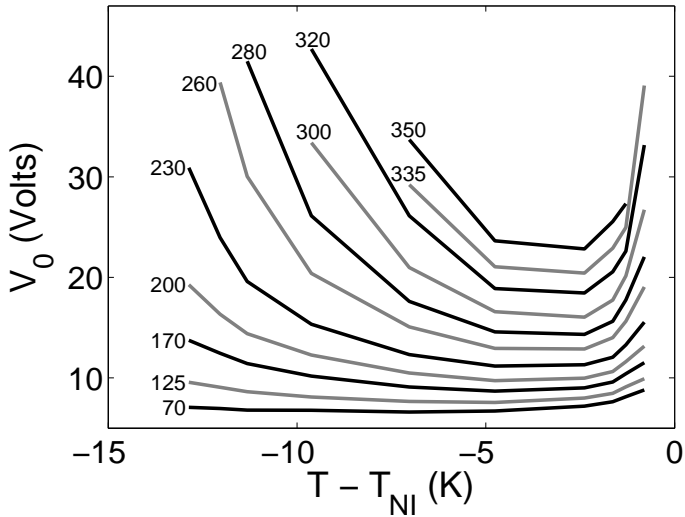}
\caption{\label{FigTempDep}Measurement of the electroconvection threshold voltage in MBBA
as a function of temperature, relative to the nematic-isotropic transition temperature $T_{NI}$,
carried out at different applied frequencies.  Each curve is
labeled with the applied frequency in Hz.  Data points are connected with lines to guide the eye.}
 \end{figure}

If the control parameter $\epsilon$ that drives pattern formation in a fluid is different at different temperatures, one may in principle create controlled spatial variation in the pattern structure using thermal inhomogeneities.  We present in Fig.~\ref{FigTempDep} measurements obtained at various applied frequencies showing the MBBA EC threshold voltage as a function of temperature.  No laser beam is present for these experiments, and the threshold voltage at a given applied frequency is simply defined as the voltage at which rolls are first observed on the sample at a fixed point that is far from the edges.  A nonmonotonic temperature dependence was also reported in our previous study \cite{GiebinkDRS04} at a single applied frequency of 70 Hz; we see from the current results that, away from the minimum in $V_0(T)$, the slope magnitude $\mid\!{dV_0/dT}\!\mid$ increases sharply with applied frequency.  These results are consistent with the very recent threshold-voltage measurements of T\'{o}th-Katona and Gleeson \cite{Toth-Katona-Gleeson04} on the nematic Mischung V (M5), which show that $\mid\!{dV_0/dT}\!\mid$ increases with frequency.  In the M5 data reported by T\'{o}th-Katona and Gleeson, $dV_0/dT$ is always negative.  T\'{o}th-Katona and Gleeson carefully noted that their EC patterns had greater spatial inhomogeneity at higher frequencies.  In the present work, as with our earlier study, we exploit this type of inhomogeneity to control the spatial variation of the pattern amplitude with an applied symmetry-breaking lateral temperature gradient.  Indeed, from the data of Fig.~\ref{FigTempDep}, one may envision two methods for controlled variation in pattern structure using a region that is warmer than its surroundings.  (1) At a bulk sample temperature such that $dV_0/dT < 0$, one can generate a locally warmed region with  $\epsilon > 0$ surrounded by a linearly stable conducting background with $\epsilon < 0$, resulting in a spatially localized pattern.  (2)  If instead $dV_0/dT > 0$ at the bulk temperature, then one can warm a local region within a \emph{preexisting} pattern to generate a local amplitude ``hole'' which (relative to the surrounding pattern) has a smaller $\epsilon$ and hence a reduced amplitude.  As discussed below, we have used laser light to successfully generate a local region with $\epsilon > 0$ and hence a localized pattern (process (1)).  We have found, however, that process (2) is much more difficult experimentally, because according to Fig.~\ref{FigTempDep}, the required bulk temperature is close to the nematic-isotropic transition temperature, and we often observed the formation of small regions of isotropic fluid that strongly distorted the pattern.  Hence in this article we will confine our attention to localized patterns produced via process (1) (\emph{i.e.} $dV_0/dT < 0$).  In passing, we should also note that if one wishes to \emph{minimize} the effects of small thermal inhomogeneities in MBBA, one should work at the temperature at which $V_0(T)$ achieves a minimum at the operating frequency.

\begin{figure}
\includegraphics[width=0.55\textwidth]{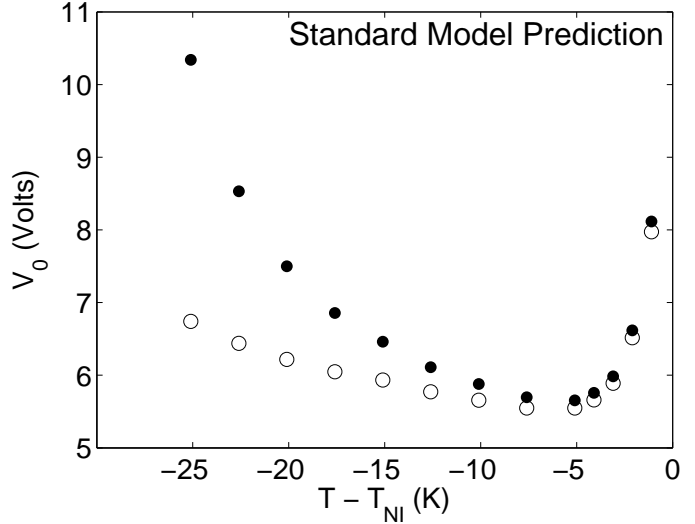}
\caption{\label{FigStandMod}Prediction of the threshold voltage at different temperatures
using the Standard Model.  The open and filled circles correspond to applied frequencies of 70~Hz and 200~Hz,
respectively.}
\end{figure}

Our experimentally observed nonmonotonic temperature dependence for the threshold voltage in MBBA can be compared to the theoretical ``Standard Model" for EC developed by Bodenschatz, Kramer, Pesch, and Zimmermann \cite{Bodenschatz88,Kramer-Pesch96,KramerAnnRevFluidMech95}.
In this model, the square of the threshold voltage $V_0$ can be expressed as stated by Kramer and Pesch \cite{Kramer-Pesch96}:

\large
\begin{equation}
\label{EqStandMod}
V_0^2= \frac {\pi^2K^{(eff)}} {         \epsilon_0\epsilon_a^{(eff)}        + \frac{\mid\alpha_2\mid \tau_q\sigma_a^{(eff)}}{\eta^{(eff)}}               }                          
\end{equation}
\normalsize
\\
where $K^{(eff)}, \epsilon_a^{(eff)}, \sigma_a^{(eff)},$ and $\eta^{(eff)}$ represent effective parameters for the elasticity, the dielectric anisotropy, the anisotropy in the electrical conductivity, and the viscosity, respectively.  For nonoblique rolls, each of these four effective parameters depends on the wavenumber $q$ selected by the pattern; furthermore, $\epsilon_a^{(eff)}$ and $\sigma_a^{(eff)}$ also depend on the frequency of the applied field.  The constants $\alpha_2$ and $\tau_q$ represent the second Leslie coefficient and the charge relaxation time, respectively.  The temperature dependence of each parameter in Eq.~\ref{EqStandMod} can be evaluated using studies from the literature \cite{Kneppe82,Gaspard73,Haller72,Diguet70}.  We normalized the literature values of the electrical conductivity to our measured room-temperature MBBA value of $\sigma_{\perp} = 8$ x $10^{-8} (\Omega \rm{m})^{-1}$ \cite{GiebinkDRS04}.  At each temperature $T$, the correct wavenumber $q$ was determined with a spreadsheet based on Eq.~\ref{EqStandMod} by adjusting $q$ until $V_0$ achieved a minimum, which provided the desired $V_0(T)$.  As an independent test of our spreadsheet algorithm, we calculated $V_0$ using as inputs the room-temperature MBBA parameters published by Bodenschatz \emph{et al.} \cite{Bodenschatz88}. Figure 3 of that paper shows $V_0$ \emph{vs.} applied frequency, based on the Standard Model with rigid boundary conditions at $25 \; \mathrm{^oC}$.  We found that our spreadsheet-calculated values for both the selected $q$ and the threshold voltage $V_0$ at ten different applied frequencies agreed with  those of Bodenschatz \emph{et al.} at a level 2\% or better.   Our spreadsheet-based Standard-Model results for $V_0$ as a function of temperature are shown in Fig.~\ref{FigStandMod} at two different applied frequencies.  The nonmonotonic temperature dependence and the strong increase in the slope \(\mid\! dV_0/dT \!\mid \) with applied frequency are both apparent in the theoretical Standard-Model prediction depicted in Fig.~\ref{FigStandMod}.  The values for $V_0$ predicted using the theory do not agree quantitatively with the experiment:  although the theory and the experimental data are consistent to about 20\% at 70 Hz, the difference is about a factor of two at 200 Hz.  This is not a serious concern, since there are four separate effective material parameters in the theory, each of which (most notably the electrical conductivity) displays considerable variation in the literature.  The tendency of $V_0(T)$ to increase at both the high and low temperature limits of the nematic phase is quite clear in the Standard Model, and we believe our experiments are consistent with this model.

\begin{figure}
\includegraphics[width=0.55\textwidth]{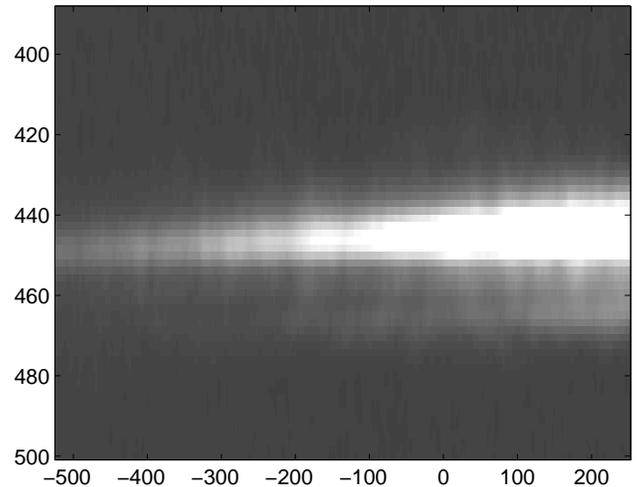}
\caption{\label{FigLaserProfile}A typical laser intensity profile.  The horizontal and vertical axes represent x and y, respectively, in units of micrometers. The width in
the y direction (FWHM) is about 11 $\mu \rm{m}$.  The total laser power
incident on the sample for the experiments was $70~\mu \rm{W}$.  A HIGHER-RESOLUTION PLOT OF THE LASER PROFILE IS AVAILABLE IN THE PUBLISHED VERSION OF THIS DOCUMENT (PRE 73, 036317, 2006).}
\end{figure}

The temperature dependence of the threshold voltage in MBBA reported in 
Fig.~\ref{FigTempDep} above confirms that it should be possible to use lateral temperature gradients in MBBA to create interesting variations in EC pattern structure, which is the basis of the experiments reported below.  These measurements are an improvement on a previous study \cite{GiebinkDRS04} in which elliptical localized EC pulses with major axes of 30 - 300 $\mu$m were created in MR-doped MBBA using a laser beam with a circular cross-section with a diameter of about 50 $\mu$m. In that work we concluded that the mechanism for the pulse creation was thermal, because essentially identical patterns were seen using anthraquinone as an alternative absorbing dye.  In the previous work, we found that the amplitude and lateral size of the laser-induced convection pattern actually decreased when the control parameter was driven past  $\epsilon_{max} \sim 0.4$.  This unexpected attenuation was attributed to the advection-driven increase in the effective thermal conductivity (or equivalently, an increase in the local Nusselt number) at larger pattern amplitudes, which quenches the temperature spike produced by the absorbed laser light.  We also found interesting roll drift in regions where the laser intensity was spatially nonuniform, but not within regions in which the laser intensity was uniform.   In the current study, we have made two experimental improvements, resulting in a simplification in the pattern structure and the dynamics.  First, we have used cylindrical lenses to stretch the laser cross-section into a long strip parallel to the director.  For a bulk sample temperature at which $dV_0/dT < 0$, a laser-light strip of this type can be used to produce a quasi-one-dimensional pattern delimited by ``soft'' \cite{CHReview} boundary conditions:  a long narrow strip in which $\epsilon(x) > 0$ is surrounded by a linearly stable state with $\epsilon < 0$.  Long narrow electroconvection channels have been generated previously using etching techniques and ``hard'' boundary conditions by Hidaka \emph{et al.} \cite{Hidaka97}, and Rasenat, Braun, and Steinberg \cite{Rasenat91, Braun91}.  Second, we have fixed the control parameter at $\epsilon_{max} = 0.1$ throughout the current study, to avoid complications caused by the amplitude dependence of the local Nusselt number.  (In an EC experiment, the constraint $\epsilon_{max} = 0.1$ maps out a curve that is displaced slightly from the usual $\epsilon_{max} = 0$ threshold curve in the plane of applied frequency \emph{vs.} applied voltage.)  To produce a nonuniform laser-intensity profile, we used a razor blade to block one-half of the length of the rectangular cross-section, so that knife-edge diffraction resulted in an inhomogeneous laser-beam profile along the director.  A typical laser intensity profile is shown in Fig.~\ref{FigLaserProfile}.  By measuring the change in the nematic-isotropic transition temperature at the point of maximum intensity within the laser-illuminated region, using both increasing and decreasing bulk temperatures, we inferred that the temperature increase at this point was about $0.9 \; \mathrm{^oC}$.  This temperature increase is not large enough to significantly change the dye absorption coefficient.

\begin{figure}
\includegraphics[width=0.55\textwidth]{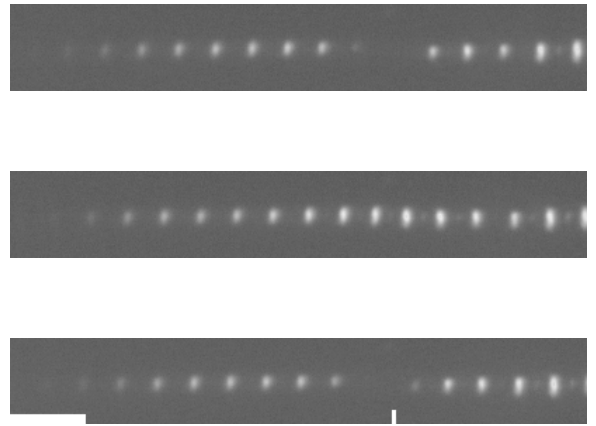}
\caption{\label{FigImage}Direct images (without background correction)
of the electroconvection patterns formed along the laser-light strip at
several different times.  From bottom to top, the images were acquired at times $t =$ 0, 9~s, and 27~s.  The applied voltage and frequency were 9.33~V and 200~Hz, respectively, which resulted in a control parameter of $\epsilon_{max} = 0.1$.  The bulk sample temperature was $T - T_{NI} = -16.7\;^{\rm o} \rm{C}$.  The white bar in the lower-left corner represents a length of $100~\mu \rm{m}$, and the short white vertical bar to the right of center in the lower panel indicates the position $x = 0$.  The image contrast has been enhanced so that the low-amplitude rolls are easier to see. }
\end{figure}

\begin{figure}
\includegraphics[width=0.55\textwidth]{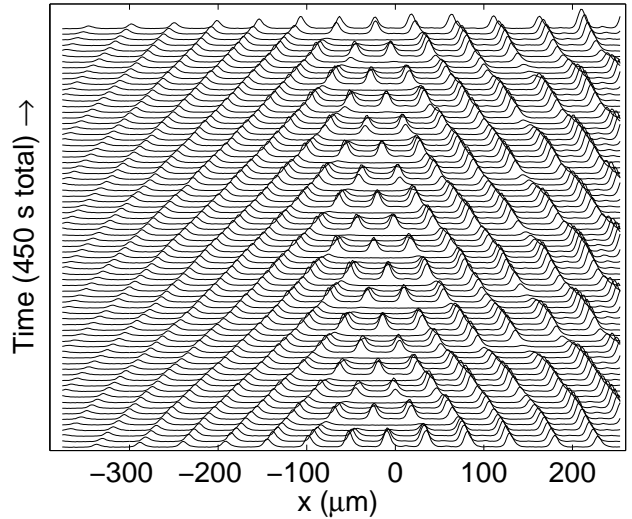}
\caption{\label{FigSpacetimeManyFrames}A spacetime plot showing the evolution of the pattern displayed in Fig.~\ref{FigImage}.  Each trace represents, at different times, the dimensionless intensity averaged over the $y$ direction.  Time evolves upward, and the total elapsed time is 450~s.}
\end{figure}

\begin{figure}
\includegraphics[width=0.55\textwidth]{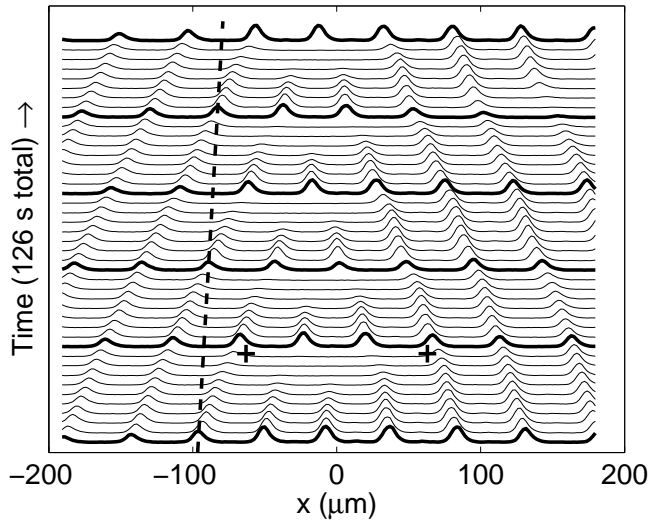}
\caption{\label{FigSpacetimeFewFrames}A spacetime plot of the pattern at higher spatial and temporal resolution.  The region referred to as the ``central zone'' in the text corresponds to the width of the amplitude gap, which is indicated (at a specific time) with the two plus signs.  Six traces showing the pattern that grows into the gap are indicated with thick lines; note that alternate thick-line traces are phase-shifted by about one-half of a cycle.  The slow overall group drift is indicated by the dashed line, which has a reciprocal slope equal to the group velocity.}
\end{figure}

Images of a localized pattern of short rolls produced along the laser strip at three different times with a control parameter of $\epsilon_{max} = 0.1$ are shown in Fig.~\ref{FigImage}.  In contrast to previous patterns commonly observed in MBBA close to threshold, in which the rolls are either stationary or drift slowly in one direction \cite{Rehberg-Ahlers91,RehbergAdvSolState89,Joets-RibottaPRL88}, the rolls localized along the laser strip show phase drifts that counterpropagate into a sink point ($x=0$), as shown with spacetime contours in Fig.~\ref{FigSpacetimeManyFrames}.  As the rolls drift, the frequency $f_{roll}$ of the amplitude variation at a specific position $x$ was found to be constant with respect to $x$ to within about 1\%.  In Fig.~\ref{FigSpacetimeFewFrames}, spacetime contours for the same images at higher spatial and temporal resolution show that in addition to counterpropagating phase drifts, there is also a strong temporally periodic amplitude modulation.  Indeed, an amplitude ``gap'' in a central region about the sink point periodically appears in which 3-4 rolls are reduced to essentially zero amplitude.   (We will subsequently refer to the region in which the amplitude gap forms as the ``central zone''.)  A new set of rolls then rapidly grows into the gap region, until the amplitude near $x = 0$  is actually higher than the amplitude outside the central zone.  Next, the central-zone rolls slowly decrease in amplitude until a gap has again formed.  This new gap is then again replaced by a set of rolls, but the new rolls are shifted in phase by about one-half cycle relative to the rolls that grew into the previous gap.  For the sample employed for Fig.~\ref{FigSpacetimeFewFrames}, we varied the applied frequency over the range $f = 160 - 430 \; \mathrm{Hz}$, while simultaneously varying the applied voltage over the range $V = 8.2 - 57 \; \mathrm{V}$ to keep the control parameter fixed at $\epsilon_{max}=0.1$.  We examined spacetime contours at fourteen different settings for $(f, V)$ and found that the same basic dynamics, with counterpropagating phase drifts and a strong amplitude modulation leading to a central-zone gap, was clearly present at each $(f, V)$ setting.  The speed of both the phase drift and the amplitude modulation was found to increase strongly with applied voltage, as we report below in detail.  In addition, this same basic counterpropagating-phase-drift/amplitude-modulation dynamic has been observed on four different independently prepared samples illuminated with a laser intensity that was nonuniform in the $x$-direction.  The dashed line in Fig.~\ref{FigSpacetimeFewFrames} shows that in addition to the counterpropagating phase drift, there is also a slow overall group drift of all the rolls toward positive $x$.  From the figure one may conclude that the group velocity for overall drift is about $1/7$ of the phase velocity corresponding to the counterpropagating roll drift.  An overall group drift is commonly observed in the literature for patterns subjected to a broken lateral symmetry \cite{Hartung91,Rehberg-Busse87,NingAhlersCannell90,Page05}.  On the other hand, the counterpropagating-roll phase drift and the temporally periodic amplitude modulation in the central zone are unexpected, and will require further elucidation.

\begin{figure}
\includegraphics[width=0.55\textwidth]{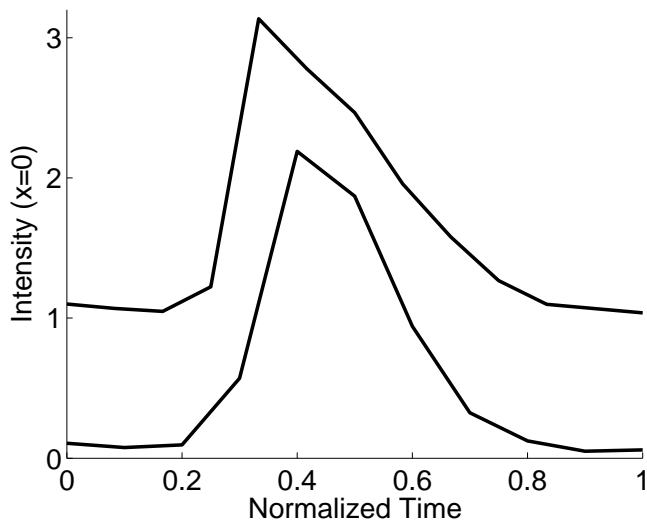}
\caption{\label{FigTimeVariation}  Measured dimensionless intensity (see Eq.~\ref{EqIntensity}) as a function of time at the position $x=0$, using two different applied voltages/frequencies.  The control parameter is $\epsilon_{max}=0.1$ for both curves.  The applied frequency and voltage is (400~Hz, 31.4~V) for the lower curve, and (200~Hz, 9.33~V) for the upper curve.  The total elapsed time along the horizontal axis is 4.3~s (lower) and 36~s (upper).  The upper curve has been offset vertically by one unit.}
\end{figure}

From Fig.~\ref{FigSpacetimeFewFrames} it is clear that the temporal dependence of the amplitude at a fixed point $x$ within the central zone is periodic but not sinusoidal.  This is shown more explicitly in Fig.~\ref{FigTimeVariation}, which displays the amplitude at $x=0$ for the data of Fig.~\ref{FigImage} as a function of time for one period at two different applied voltages.  The rising edge is considerably faster than the falling edge.
This type of asymmetric temporal behavior, which is usually referred to as relaxation-oscillation in the nonlinear dynamics literature \cite{StrogatzBook94}, is fundamentally nonlinear, since linear instability analysis will always result in a (in general complex) exponential time dependence.

\begin{figure}
\includegraphics[width=0.55\textwidth]{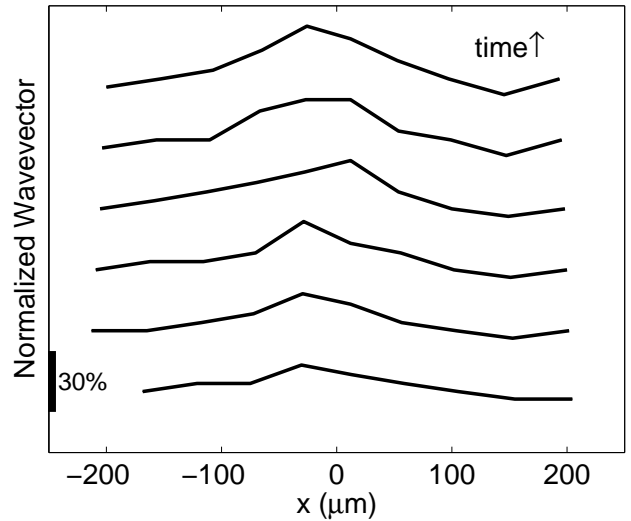}
\caption{\label{FigWavenumber}Variation of the wavenumber with position along the x-axis.  The data employed are the earliest six traces from the spacetime plot of Fig.~\ref{FigSpacetimeFewFrames}, before the first amplitude gap forms.  The wavenumber is normalized to its value on the far left.  Time evolves upward, with an interval of 3~s between consecutive plots.  The thick vertical bar on the lower part of the left axis represents a wavenumber variation of 30\%; thus, we observe on this figure that the peak normalized wavenumber near $x = 0$ increases (with time) from about 1.1 to about 1.3. }
\end{figure}

As the phase-drift and amplitude-modulation processes occur, the wavenumber profile $q(x)$ of the pattern varies with time.  The variation is shown in Fig.~\ref{FigWavenumber}, where $q(x)$, normalized to its value for the left-most spatial period in the pattern, is shown at several different times before the amplitude gap forms.  At a given time, the  wavenumber $q(x)$ has its maximum near $x=0$, and is nearly symmetric about that maximum.  The maximum fractional difference between the wavenumber near the left edge of the strip and the wavenumber at $x=0$ is about 30\%.  This exceeds the Eckhaus bandwidth for stable periodic states at a control parameter of $\epsilon = 0.1$, which is $\sqrt{\epsilon/3} = 18 \%$ \cite{Lowe-Gollub-Eckhaus85}.  We may therefore anticipate that wavelength-changing processes driven by the Eckhaus instability are potentially relevant in our experiments.  Our observation in Fig.~\ref{FigSpacetimeFewFrames} that the location of the amplitude the gap is selected by the system in a central region about $x = 0$ (i.e., far from the boundaries) appears to be consistent with predictions of Kramer and Zimmermann \cite{KramerZimmermann85} regarding the location of the reduced-amplitude wavelength-changing zone within an Eckhaus-destabilized pattern.  These authors argued that, while the location of the wavelength-changing process in an infinite system is arbitrary, in a real system this process will occur in a centralized ``weak" region far from the boundaries, in a manner consistent with Taylor vortex flow experiments \cite{Cannell-Ahlers83} and simulations \cite{Dominguez-Lerma86,Ahlers89}.

It is also of interest to compare the roll-dynamic timescales that we observe experimentally to the relevant Giznburg-Landau (GL) timescale. The one-dimensional GL equation for a slowly-varying amplitude $A(x, t)$ in the weakly nonlinear regime ($\epsilon << 1$) can be written \cite{CHReview, Ahlers89}

\begin{figure}
\includegraphics[width=0.55\textwidth]{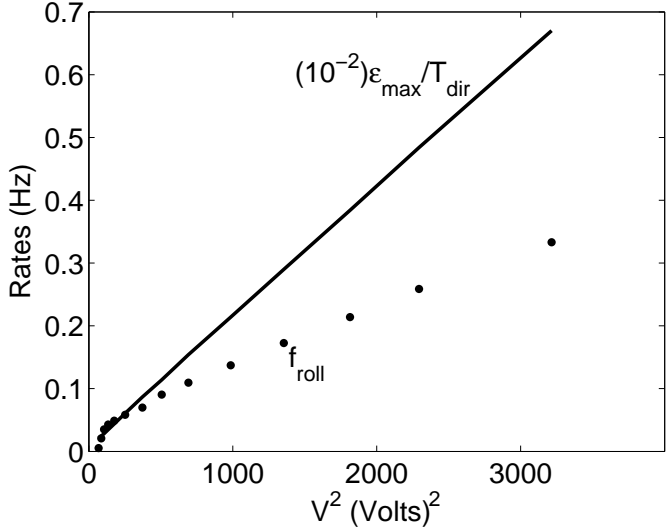}
\caption{\label{FigRates}The individual points are the measured roll frequencies $f_{roll}$ at the position $x=0$ shown as a function of the square of the applied voltage $V$.  The applied frequency and voltage were both varied so that the  control parameter $\epsilon_{max}$ is equal to 0.1 at each individual point displayed.  The solid line is the director relaxation rate $\epsilon_{max}/T_{dir}$, scaled by a factor of $10^{-2}$ so that $f_{roll}$ and $\epsilon_{max}/T_{dir}$ can be displayed on the same graph.}
\end{figure}

\large
\begin{equation}
\label{EqGL}
\tau \, \partial_tA(x,t) = \xi_0^2 \, \partial_x^2A(x,t) + \epsilon \, A(x,t) - g_0\mid A \mid^2A(x,t)                        
\end{equation}
\normalsize
\\
in which the space and time scales are represented by $\xi_0$ and $\tau$, respectively, with $\epsilon$ serving as the control parameter and $g_0$ as the usual negative-feedback coefficient.  For electroconvection, $\tau$ is given approximately by the director relaxation time $T_{dir}$, since this is generally the longest timescale appearing in the microscopic equations of motion \cite{Bodenschatz88, Kramer-Pesch96}.  It is possible that the control parameter $\epsilon$ can vary on position and time scales that are large compared to $\xi_0$ and $\tau$, respectively \cite{Malomed94}.  To compare the GL timescale to our experimentally measured timescales, we designed an experiment in which we measured the temporal frequency $f_{roll}$ of the intensity variation at a fixed point $x$ as a function of the applied voltage across the sample.   In this measurement, it is again important to note that the applied voltage and frequency were varied together in a manner such that the control parameter was held constant at $\epsilon_{max} = 0.1$.  Results for the measured frequency $f_{roll}$ \emph{vs.} the applied voltage across the sample are shown in Fig.~\ref{FigRates}, with $f_{roll}$ measured at $x=0$.  By using the Standard Model to compute $T_{dir}$ at each applied voltage, we may also show $\epsilon_{max}/T_{dir}$, scaled down by a factor of $10^{-2}$, on the same graph.  Clearly, at a given applied voltage, the measured frequency $f_{roll}$ is always much smaller than $\epsilon_{max}/T_{dir}$.  Thus the temporal variations in amplitude and phase that we observe is a slowly varying temporal modulation vis-\'{a}-vis the GL equation.
\\

\section{Conclusions and Future Work}

In considering potential theoretical models to explain our experiments, we consider first the theory of very small parameter ramps in pattern-forming media due to Kramer and coworkers \cite{Kramer82}.  A ramp within a one-dimensional pattern selects a specific (though in general nonuniversal) wavenumber profile function $q(x)$.  It is well known that drift can occur in ramped medium.  Riecke and Paap \cite{Riecke87}, for example, showed that if $q(x)$ penetrates into the Eckhaus-unstable band, phase drift may occur.  In addition, Rehberg \emph{et al.} \cite{Rehberg-Busse87} carried out experiments in which two ramps with opposite slope at the ends of a Rayleigh-B\'{e}rnard cell selected two different wavenumber profiles $q_{left}(x)$ and $q_{right}(x)$, which can generate a unidirectional phase drift from high $q$ toward low $q$.   The authors explained these experiments with a qualitative model showing that drift will occur whenever there is no overlap between the finite-width selection bands centered on $q_{left}(x)$ and $q_{right}(x)$.  For our experiments, it is reasonable to conclude that the slow uniform \emph{group drift} results from the thermal ramp.  However, the observed counterpropagating phase drift and the central-zone amplitude modulation do not appear to be compatible with previously reported drift mechanisms in ramped media for two reasons.  First, although our ramp, in the form of an imposed thermal gradient, has the same sign everywhere in the pattern, the observed phase drift is counterpropagating rather than unidirectional.  Second, in addition to phase drift, we observe a strong periodic amplitude modulation in the region close to $x=0$.  The incompatibility of our results with ramp-theory drifts is perhaps not surprising in view of the requirement of very small ramps in the theory.

To the best of our knowledge, all theoretical work on models of electroconvection has been carried out assuming a uniform isothermal medium for the spatially extended rolls.   Thus, a theoretical model of electroconvection patterns subjected to a local temperature gradient, and in particular the use of local heating to produce \emph{localized} pattern formation with soft boundary conditions, has not yet been formulated.  Based on the results above, we believe that, in closing, we can offer suggestions on potentially useful avenues for theoretical development on these interesting dynamics.  In developing a new model, one must demand explanations for the following salient aspects of the data, which represent distinct new features vis-\'a-vis traditional spatially extended EC rolls in uniform MBBA:

\begin{itemize}
\item opposing phase drifts for very negative and very positive $x$, resulting in counterpropagating rolls;
\item a strong amplitude modulation resulting in the temporally periodic formation of a ``gap"  in the central region, with a temporal profile that is periodic but not sinusoidal.

\end{itemize}

We wish to point out two theoretical approaches that could prove useful in developing a model for these experimental features.

(1) First, it is clear from Fig.~\ref{FigWavenumber} that the wavenumber bandwidth is driven back and forth across the Eckhaus boundary.  Therefore, if one wishes to apply solutions to the GL equation applicable to our experiments, one should include not only the usual constant-amplitude solutions $A(x,t) = g_0^{-1/2}\sqrt{ \epsilon - k^2  \xi_0^2  } \; e^{ikx}$, where $k = q - q_0$, but also the amplitude-modulated saddle-point (SP) solutions that separate basins of attraction of the constant-amplitude solutions.  These SP solutions, which drive the wavelength-changing processes that arise when the Eckhaus boundary is crossed, were considered in detail in two widely cited papers by Kramer, Schober, and Zimmerman \cite{KramerZimmermann85, KramerSchoberZimmermann88}.  The simplest SP solution considered by these authors is the function

\large
\begin{equation}
\label{EqKZ}
A_{SP}(x,t) = \frac{e^{ikx} }{\sqrt{g_0}}     \left[\sqrt{2}  \: k\xi_0 +  i\delta(t)\tanh(\frac{\delta(t) \: x}{\sqrt{2} \: \xi_0}  )\right]             
\end{equation}
\normalsize
\\
where $\delta(t) = \sqrt{ \epsilon(t) -3k^2\xi_0^2 } \;$  is real, and we have included a potentially slowly-varying control parameter $\epsilon(t)$.  Kramer and Zimmermann \cite{KramerZimmermann85} pointed out that for large $\mid\! x \!\mid$, $A_{SP}$ takes the form of the usual periodic GL solution (proportional to $e^{ikx}$); however, due to the tanh function, there is a phase shift between the periodic solutions at $x= \pm \, \infty $.  In view of the experiments, an interesting characteristic of Eq.~\ref{EqKZ} is the fact that if the control parameter is changed slowly, the phase of $A_{SP}$ will change in opposite directions at $x= \pm \, \infty $.

2.  Since counterpropagating traveling waves are seen under all experimental conditions explored, it is certainly reasonable to consider application of two coupled complex Ginzburg-Landau equations (CGLE)\cite{vanSaarloosTutorial}, in which the parameters $\xi_0^2$ and $g_0$ in Eq. \ref{EqGL} above will have both real and imaginary parts.  A direct comparison of experiment with CGLE-based theory can be difficult, because the CGLE can describe an extraordinary variety of different traveling-wave dynamics, and the coefficients of the equation are in general difficult to determine experimentally\cite{Pastur03}.  However, it is well-known that a reduction of the dimensionality or localization in the experimental patterns, which have both been achieved in the present experiments by using a long strip of laser light to localize the rolls, can reduce considerably the difficulty of determining the coefficients.  A useful starting point would appear to be the CGLE modeling approach adopted by Pastur \emph{et al.}\cite{Pastur03} in describing their heated-wire experiments.  These authors observed left- and right-traveling waves separated by sources and sinks.  If the CGLE model employed by Pastur \emph{et al.} could be extended to the case of nonuniform spatial heating by introducing a spatially dependent control parameter $\epsilon(x)$, it would be very interesting to see if the traveling-wave dynamics resulting from the model displayed characteristics that reproduced the experimental results above.

To conclude, we have presented experimental results on electroconvection-pattern formation in a narrow strip in which the lateral symmetry has been broken with a laser-induced thermal gradient.  Our main purpose was to report new experimental results that differ in a qualitative manner from previous experiments on broken-symmetry fluidized patterns.  Our measured temperature dependence of the threshold voltage showed that is possible to create a thermally localized supercritical strip.  The spacetime profiles along the strip displayed a slow counterpropagating phase drift and an asymmetric relaxation-oscillation temporal profile, in addition to a strong slowly varying temporally periodic amplitude modulation that generates a central-zone gap.  Measurement of the spatial variation of the wavenumber displayed a wavenumber bandwidth  that exceeded the Eckhaus stability limit.  We noted some simple connections to Eckhaus unstable states and to the complex Ginzburg-Landau equation in an attempt to provide some potential guidance for the development of a theoretical model to explain the results.

We are at present upgrading our apparatus and software to gain an improved understanding of the patterns discussed above.  There are two future tasks that we believe are especially exigent.  First, it is very important to carry out a \emph{direct} measurement of the temperature as a function of position along the pattern strip.  This would allow a determination of the control parameter $\epsilon(x)$ as a function of position, which is important for the development of a model.  This is difficult experimentally since the maximum temperature variation is only on the order of $1 \; \mathrm{^oC}$, but is crucial nonetheless.  Second, one should investigate whether numerical simulations of coupled CGL equations with a driving force consistent with Fig.~\ref{FigLaserProfile} can show an amplitude modulation and counterpropagating phase drifts akin to the experimental results.  Code-writing for such simulations is currently in progress.

\begin{acknowledgments}
We gratefully acknowledge very helpful conversations with Guenter Ahlers, Bill Collins, Stephen Morris, Matthias Schr\"{o}ter, Harry Swinney, and Richard Wiener.  DRS is very grateful to Ingo Rehberg and Werner Pesch for their kind hospitality and important feedback on experimental and modeling questions during a brief visit to Bayreuth.  This work was supported in part by a grant from the National Science Foundation.  Acknowledgment is also made to the donors of the Petroleum Research Fund, administered by the American Chemical Society, for partial support of this research.

\end{acknowledgments}

\section{References}


\end{document}